\documentclass[letterpaper, prd, aps, twocolumn, preprintnumbers, floatfix, nofootinbib]{revtex4}

\usepackage[dvips]{graphicx}
\usepackage{epsf}
\usepackage{amsmath}
\usepackage{amssymb}
\usepackage[colorlinks=true,urlcolor=blue,linkcolor=blue,citecolor=blue]{hyperref}
\usepackage{enumerate}

\def\be{\begin{equation}}
\def\ee{\end{equation}}
\def\bea{\begin{eqnarray}}
\def\eea{\end{eqnarray}}

\begin{document}

\title{Cosmic Microwave Background Spectral Distortions from Cosmic String Loops}

 \author{Madeleine Anthonisen${}^1$%
   \footnote{\href{mailto:madeleine.anthonisen@mail.mcgill.ca}
     {madeleine.anthonisen@mail.mcgill.ca}},
   Robert Brandenberger${}^{1,\,2}$%
   \footnote{\href{mailto:rhb@physics.mcgill.ca}{rhb@physics.mcgill.ca}},
   Alex Lagu\"e${}^1$%
   \footnote{\href{matilto:alex.lague@mail.mcgill.ca}{alex.lague@mail.mcgill.ca}},
   Ian A.~Morrison${}^1$%
   \footnote{\href{mailto:imorrison@physics.mcgill.ca}{imorrison@physics.mcgill.ca}},
   and Daixi Xia${}^1$%
   \footnote{\href{mailto:daixi.xia@mail.mcgill.ca}{daixi.xia@mail.mcgill.ca}}}

\affiliation{\vspace{12pt}${}^1$Department of Physics, McGill University, 
  Montr\'eal, QC, H3A 2T8, Canada \\
${}^2$Institute for Theoretical Studies, ETH Z\"urich, 
  CH-8092 Z\"urich, Switzerland }


\begin{abstract}

Cosmic string loops contain cusps which decay by emitting bursts of 
particles. A significant fraction of the released energy is in the 
form of photons. These photons are injected non-thermally and can 
hence cause spectral distortions of the Cosmic Microwave Background 
(CMB). Under the assumption that cusps are robust against 
gravitational back-reaction, we compute the fractional energy 
density released as photons in the redshift interval where such 
non-thermal photon injection causes CMB spectral distortions. 
Whereas current constraints on such spectral distortions are not
strong enough to constrain the string tension, future missions 
such as the PIXIE experiment will be able to provide limits which rule 
out a range of string tensions between $G \mu \sim 10^{-15}$ and
$G \mu \sim 10^{-12}$, thus ruling out particle physics models yielding 
these kind of intermediate-scale cosmic strings.

\end{abstract}

\maketitle

\section{Introduction}

Cosmic strings \cite{Kibble1} are topologically stable solutions of 
the classical field equations in many particle physics models beyond 
the {\it Standard Model}. If matter is described by such a model, 
then a simple causality argument of Kibble \cite{Kibble2} ensures 
that a network of strings will form during a symmetry-breaking phase 
transition in the early universe and will persist to the present time. 
Both in the 1980s (see e.g. \cite{Vil, HK, RHB1} for reviews) and more 
recently (see e.g. \cite{RHB2} for a recent review) there has been a 
lot of work devoted to the observational signatures of such strings. 
The network of cosmic strings is described by a single free parameter, 
namely the mass per unit length $\mu$ of a string, which is related to 
the energy scale $\eta$ of the symmetry-breaking phase transition via 
$\mu = d \eta^2$, where $d$ is a number of order unity (which, in 
particular, to a first approximation does not depend on coupling 
constants in the particle physics model).

Cosmic strings carry energy and hence their gravitational effects 
can lead to cosmological signatures. In fact, at one point, it was 
conjectured \cite{CSold}  that cosmic strings could seed all of the 
observed structure in the Universe. With the discovery of acoustic 
oscillations in the angular power spectrum of the Cosmic Microwave 
Background (CMB) \cite{Boomerang} this possibility was ruled out. 
Currently, the best and most solid constraints on the string tension 
come from precision measurements of the CMB angular power spectrum. 
They are \cite{Dvorkin, Planck} 
\be \label{bound}
G \mu \, <  \, 2 \times 10^{-7}
\ee 
(see \cite{older} for older results), where $G$ is Newton's gravitational 
constant (the string tension is typically expressed in terms of a 
dimensionless number by multiplying by $G$).\footnote{We
will use natural units throughout this note.} This already rules out a 
set of particle physics models which have a symmetry breaking scale at 
the high end of the ``Grand Unification'' class. Tightening these 
constraints or providing new constraints is of great interest since it 
will provide new ways to test particle physics beyond the Standard Model, 
ways which are complementary to the test at accelerators such as the 
Large Hadron Collider (LHC) (see \cite{RHB3} for an elaboration on this theme).

Cosmic strings also form closed string loops.
Kibble's causality arguments coupled to studies of the scaling solutions 
of the Boltzmann-type equation which describes the time evolution of the 
network of strings show \cite{Vil, HK, RHB1} that there will at all times 
be a scaling network of infinite strings with typical separation and 
correlation length of the order $t$ (the cosmic time), as well as a distribution of 
string loops which are the result of interactions of the long strings. 
This distribution of string loops is statistically independent of 
time if all lengths are scaled to the Hubble radius $t$.

There has been a lot of work on the gravitational effects of 
cosmic strings (see e.g. \cite{RHB2} for a recent review). The long 
strings lead to distinctive line discontinuities in CMB temperature maps 
\cite{KS}, to patches in the sky with direct B-mode polarization 
\cite{Holder1} and to wedges in 21cm redshift maps at high redshifts 
with extra absorption \cite{Holder2}. Cosmic string loops may form the 
seeds of globular clusters \cite{Lin}, they may assist the formation of 
high redshift super-massive black holes \cite{SMBH}, and they may lead to 
ultra-compact mini-halos embedded within galaxies \cite{Maddy}. All of 
these effects are due to the gravity of cosmic strings.

In this note we study a non-gravitational effect of cosmic string loops.
As argued in \cite{KT} and recently confirmed numerically 
in \cite{Olum}, cosmic string loops are not smooth but generically 
contain {\it cusps}. 
Cusps are unstable to decay and will evaporate into jets of 
particles.
A significant fraction of the energy lost in this process
goes into non-thermal photons. 
Non-thermal photons generated during in the redshift range
$3 \times 10^6 > z > 3.6 \times 10^3$ (where $z$ denotes cosmological redshift)
are not able to reach equilibrium with the CMB.
As a result, these photons cause spectral distortions of the CMB 
which can be measured by future experiments. 
We show below that, at least for a class of cosmic string models, 
results from the planned PIXIE experiment could 
rule out string loops with string tensions in the
range $10^{-15} \lesssim G \mu \lesssim 10^{12}$.


Consequences of cusp annihilation for cosmology have been studied earlier
in the context of ultra-high energy cosmic ray and neutrino production
from strings \cite{UHECR}.
These works did not consider spectral distortions. Spectral distortions
caused by strings were considered in \cite{Carter} in the context of
decaying cosmic defects, in \cite{Tashiro1} in the context of acoustic waves
produced by cosmic strings, and in \cite{Tashiro2} in the context of
superconducting cosmic strings.

Effects of the final collapse of a string loop
(once the radius becomes comparable to the width) on black hole formation
\cite{CSBH} and on dark matter production \cite{Rachel} have also been studied.
Note that the photon production during the final loop collapse is negligible
in magnitude compared to the amount of energy released from string cusps, and
hence we do not consider this final collapse in this paper.

\section{Energy Loss from Cusp Annihilation and Gravitational Radiation}

We will work in the context of a simple one-scale model
for the cosmic string loop distribution according to which loops are 
formed at times $t$ with a radius
\be \label{birth}
R_i(t) \, = \,\frac{\alpha}{\beta} t \, ,
\ee
where $\alpha$ and $\beta$ are constants. The average length of a string 
loop is $l = \beta R$, and the constant $\alpha$ 
relates the string length at the time of formation to the time. 
Analytic arguments suggest $\beta \sim 10$, 
while numerical cosmic string evolution 
simulations \cite{CSsimuls} give the rough estimate $\alpha \sim 0.1$.
Within a radiation-dominated epoch the resulting number 
density of string loops per unit radius is given by
\bea \label{dist1}
n(R, t) \, &=& \, N \alpha^{5/2} \beta^{-5/2} t^{-3/2} R^{-5/2} ,
\nonumber \\
& & \text{for } R_c(t) < R < \frac{\alpha}{\beta} t \, .
\eea
Here the constant $N$ is determined by the number of long string 
segments for Hubble volume. Numerical simulations 
indicate that this constant is of the order $N \sim 10$ 
\cite{CSsimuls}. The origin of the lower bound $R_c(t)$
will be explained momentarily.

Once formed, loops slowly decay 
through numerous channels,\footnote{%
  The defects we consider in this paper are topologically stable.
} 
and the number density (\ref{dist1}) does not include these effects.
As a simple way to incorporate these decay processes into $n(R,t)$, we 
impose a lower cutoff $R_c(t)$ on the radius in the distribution of 
string loops.
This cutoff is the radius for which a string loop will evaporate
completely within one Hubble time scale. Below the cutoff the number density
is roughly:
\be \label{dist2}
n(R, t) \, \sim \, N \alpha^{5/2} \beta^{-5/2} t^{-3/2} (R_c(t))^{-5/2}, 
\quad R < R_c  .
\ee
Within our model there are two competing
decay mechanisms: decay through the emission of gravitational 
radiation \cite{grav}, and decay through cusp annihilation.
The dominant mechanism determines the cutoff scale $R_c(t)$.

The emission of gravitational radiation is independent of time
as well as the radius of the loop and satisfies the simple relation
\be
{\dot R} \, = - \, \gamma G \mu  ,
\ee
where the dot denotes a derivative with respect to $t$ and
where $\gamma$ is another constant which is related to the strength of 
gravitational radiation and which must be determined from numerical 
simulations. Its value is of the order $\gamma \sim 10$. 
Noting that the energy contained in a loop is
$E = \mu l = \mu \beta R$, the power emitted by a loop due to 
gravitational waves is thus
\be \label{GWrad}
\frac{dE}{dt} \bigg|_{{\rm GW}} \, = \, \beta \gamma G \mu^2 \, .
\ee
The critical radius may then be estimated by
$R_c = - \dot{R} t$ which yields
\be
R^{\rm GW}_c(t) \, = \gamma G \mu t \, .
\ee

Next we analyze loop decay through cusp annihilation.
When speaking about string cusps, one is working in the {\it Nambu-Goto} 
approximation of cosmic string dynamics in which the finite thickness of 
the string is neglected. This is an excellent approximation in the context 
of cosmology since the spatial extent of the string is cosmological whereas 
the width is microphysical. However, at a string cusp the Nambu-Goto 
approximation breaks down. If we consider a string loop with characteristic 
length scale $R$, then, as shown in \cite{cusp}, the two string segments of 
the loop around the cusp will overlap (i.e. their distance will be smaller 
than the string width) over a distance 
\be
l(R) \,=\, \kappa R^{2/3} w^{1/3} \, ,
\ee
where $\kappa$ is an order unity constant and
$w$ is the string width. This width is related to the string
tension by $w \sim \eta^{-1} \sim \mu^{-1/2}$.
This region contains an energy of
\be
\delta E \, = \, 2 \mu l(R) \, = \, 2 \kappa \mu R^{2/3} w^{1/3} \, .
\ee

Locally, a cusp looks like an overlapping string and
antistring segment. No topology protects this
region of the string from exploding into a burst
of particles. The primary particles in this burst
will consist of quanta of the scalar and gauge fields
which make up the string. The corresponding
particles are unstable and will decay into a jet
of stable Standard Model particles. We call this
the {\it cusp annihilation} process \cite{RHB4}. Most of
the energy will end up in pions, neutrinos and
photons. We expect that a fraction $f$ of order
one will end up in photons. Since the strings are
out-of-equilibrium field configurations, the photons
resulting from cusp annihilation will be very energetic
and out of thermal equilibrium. Photons emitted
in the redshift range $3 \times 10^6 > z > 3.6\times 10^3$
will not be able to thermalize with the CMB and will
hence lead to CMB spectral distortions. Observational
limits of CMB spectral distortions hence can lead
to limits on the cosmic string model.

Since there is of the order one cusp per loop oscillation
time, and since the loop oscillation time is of the
order of the radius $R$, the rate of energy loss
of a string loop due to cusp annihilation is
\be \label{cusprad}
\frac{\Delta E}{\Delta t} \bigg|_{{\rm{cusp}}} \, 
= \, 2 \kappa \mu \left(\frac{w}{R}\right)^{1/3} .
\ee
It follows that the rate of change in the loop radius due to cusp
annihilation is
\be
\dot{R} \,=\, - \frac{2 \kappa }{\beta} \left(\frac{w}{R}\right)^{1/3} .
\ee
The decay timescale for a loop of initial radius $R_i$ may be obtained
from the approximation
\be
R(t) \,\approx\, t \dot{R}(t_i) + R_i .
\ee
The resulting cutoff in the loop distribution is
\be
R_c^{\rm cusp}(t) \,=\, 
\left(\frac{2\kappa}{\beta}\right)^{3/4} w^{1/4} t^{3/4} .
\ee

We may now compare the cutoff radii due to gravitational radiation
and cusp annihilation. These cutoffs scale with time like
$R_c^{\rm GW}(t) \sim t$ and $R_c^{\rm cusp}(t) \sim t^{3/4}$ respectively.
Thus, at early times cusp annihilation provides the larger cutoff,
while at sufficiently late times the cutoff due to gravitational radiation 
is larger.
The cutoffs are equal at the ``cross-over'' time $T$
\be
T \,=\, \frac{8 \kappa^3 w}{\beta^3 \gamma^4\mu^4 G^4} 
 = \frac{8 \kappa^3 G^{1/2}}{\beta^3 \gamma^4 \lambda^{1/2} } 
 \frac{1}{(\mu G)^{9/2}} .
\ee
In the last equality we have made the relationship between $w$ and
$\mu$ precise by inserting $w = \lambda^{-1/2} \mu^{-1/2}$, where $\lambda$
is a model-dependent Higgs coupling constant which depends on, e.g., 
the coupling constant of the QFT generating the cosmic strings.
The cross-over time may also be described in terms of the power
emission as follows: the power emitted through gravitational
radiation and cusp annihilation, i.e.~(\ref{GWrad}) and (\ref{cusprad}),
become equal at a radius
\be
  R_{\star} \,=\, \frac{8\kappa^3  G^{1/2}}
  {\gamma^3 \beta^3 \lambda^{1/2} (\mu G)^{7/2}} .
\ee
For $R > R_{\star}$ gravitational radiation is more efficient, while
for $R < R_{\star}$ cusp annihilation is more efficient. The time $T$
is the time when $R_c^{\rm GW}(T) = R_c^{\rm cusp}(T) = R_\star$.

To summarize, we take as our cutoff of the loop number density 
distribution $n(R,t)$
\bea
R_c(t) &=& R_c^{\rm cusp}(t) \Theta(T-t) + R_c^{\rm GW}(t) \Theta(t-T) 
\nonumber \\
&=& \gamma \mu G \left( T^{1/4} t^{3/4}\Theta(T-t) + t \Theta(t-T) \right) ,
\label{Rc}
\eea
where as usual $\Theta(x)$ denotes the Heaviside function.
As we will see in the next section, the energy
density in photons produced from cusp annihilation is
dominated by loops with radius near the cutoff radius.

\section{Calculation of Photon Production from Cusps}

We now turn to the calculation of the photon energy density
input from cusp annihilation. We assume that a fraction $f$
of the cusp energy ends up in non-thermal photons. Photons
released at redshifts larger than $z_{in} \sim 3 \times 10^6$
are able to thermalize. Hence, we are interested in redshifts
between $z_{in}$ and the the redshift of re-ionization. 
To slightly simplify the algebra, we will in fact focus on redshifts
larger than the redshift of matter radiation equality which is
$z_{eq} = 3.6 \times 10^3$. The corresponding times are
$t_{in}$ and $t_{eq}$. We will focus on some time $t$ in
the window $t_{in} < t < t_{eq}$.

All cusps present between $t_{in}$ and $t$ will contribute
to the non-thermal energy in photons. At each time $t^{\prime}$
between $t_{in}$ and $t$, all loops present at that time
will radiate photons via cusp annihilations. The energy
density in photons produced at time $t^{\prime}$ redshifts like
radiation between $t^{\prime}$ and $t$. Since there is roughly
one cusp for loop oscillation time $R$, the non-thermal
photon energy density is
\bea \label{rhogeneral}
\rho_{\gamma}(t) \, &=& \, \int_{t_{in}}^{t} dt^{\prime} \left( \frac{a(t^{\prime})}{a(t)} \right)^4 \\
& & \times \int_0^{\alpha t^{\prime}/\beta} dR n(R, t^{\prime}) \frac{1}{R} f 
2 \kappa \mu R^{2/3} w^{1/3} \, . 
\nonumber
\eea
Recall that our calculations are in the radiation-dominated phase,
so $a(t) \propto  t^{1/2}$.
It will be convenient later to have this quantity normalized
with respect to the background (photon) energy density $\rho_b(t)$.
From the Friedmann equation we easily compute
\be
\rho_b(t) \, = \, \frac{3}{32\pi} \frac{1}{G t^2} \, .
\ee
Combining these equations we have
\bea \label{rhofraction}
  \frac{\rho_\gamma(t)}{\rho_b(t)}
  &=& \frac{64\pi}{3} f \kappa G \mu w^{1/3}
  \nonumber \\ & & \times
  \int_{t_{in}}^t dt' (t')^2
  \int_0^{\alpha t'/\beta} dR \, n(R,t') R^{-1/3} . \quad
\eea

The integral over $R$ may be approximated by
\bea \label{Rintegral}
& & \int_0^{\alpha t'/\beta} dR \, n(R,t') R^{-1/3} 
\nonumber \\ & & \quad
 \approx N \left(\frac{\alpha}{\beta} \right)^{5/2}
 (t')^{-3/2} \frac{6}{11} R_c(t')^{-11/6} .
\eea
We obtain this by considering the integral over the range
$R_c(t) \le R < \alpha t/\beta$. 
In this region the loop distribution $n(R,t)$ 
behaves like $R^{-5/2}$, and as a result the integral is peaked
near $R\sim R_c(t)$. There is also a contribution from the integral
over the region $0 \le R \le R_c(t)$ which is, at most, of the same
order of magnitude. Since the precise form of the loop distribution
in this region is not known, we drop this contribution
with the understanding that our calculation is accurate only to within an
order of magnitude. Inserting (\ref{Rintegral}) into (\ref{rhofraction})
and tidying up we obtain
\be \label{halfway}
  \frac{\rho_\gamma(t)}{\rho_b(t)}
  = C_1 G^{1/6} (\mu G)^{5/6}
  \int_{t_{in}}^t dt' (t')^{1/2} R_c(t')^{-11/6}  \, ,\,
\ee
where we have defined the constant
\be
  C_1 = \frac{128\pi}{11} \left(\frac{\alpha}{\beta}\right)^{5/2} 
  N f \kappa \lambda^{-1/6} .
\ee

The remaining integral over $t'$ may be performed exactly.
For our purposes we are interested in maximizing the fraction
$\rho_\gamma(t)/\rho_b(t)$. Since the integrand in (\ref{halfway})
is positive, we achieve the maximum value by letting $t \to t_{eq}$.
It is also convenient to recast factors of the string tension $\mu G$ 
in terms of the cross-over time $T$ (recall $T \propto (\mu G)^{-9/2}$).
Inserting our expression (\ref{Rc}) for the cutoff radius
and tidying up further we obtain
\bea
  \frac{\rho_\gamma(t_{eq})}{\rho_b(t_{eq})}
  &=& C_2 G^{1/18} T ^{2/9} 
  \nonumber \\ & &
  \times \int_{t_{in}}^{t_{eq}} dt'
    \Big[ T^{-11/24} (t')^{-7/8} \Theta(T-t') 
    \nonumber \\ & & \phantom{ \times \int_{t_{in}}^{t_{eq}} dt' \Big[}
    + (t')^{-4/3} \Theta(t'-T)
    \big] , \;\quad 
\eea
where the overall constant is now
\be
  C_2 = C_1 \gamma^{-11/6}
  \left(\frac{8 \kappa^3 }{\gamma^4\beta^3 \lambda^{1/2}} \right)^{-2/9} .
\ee
There are now three possible cases, depending on whether
the cross-over time $T$ lies below, within, or above the range of
integration.

{\it Case (i):} $T \le t_{in}$. For this case 
$R_c(t) = R_c^{\rm GW}(t)$ over the entire range of integration.
After integrating we obtain
\be
  \frac{\rho_\gamma(t_{eq})}{\rho_b(t_{eq})}
  = 3 C_2 G^{1/18} T^{2/9} \left[ (t_{in})^{-1/3} - (t_{eq})^{-1/3}\right] .
\ee
Note that this scales like $T^{2/9}$, and so is maximized by letting
$T \to t_{in}$.

{\it Case (ii):} $t_{eq} \le T$. For this case $R_c(t) = R_c^{\rm cusp}(t)$
over the entire range of integration. Integrating yields the
expression
\be
  \frac{\rho_\gamma(t_{eq})}{\rho_b(t_{eq})}
  = 8 C_2 G^{1/18} T^{-17/72} \left[ (t_{eq})^{1/8} - (t_{in})^{1/8}\right] .
\ee
This expression scales like $T^{-17/72}$ and so is maximized by 
letting $T \to t_{eq}$.

{\it Case (iii):} $t_{in} \le T \le t_{eq}$. 
For this case the functional form of the cutoff radius $R_c(t)$ changes
within the integration region. The result of integration is
\bea
  \frac{\rho_\gamma(t_{eq})}{\rho_b(t_{eq})} 
  &=& 
  C_2 G^{1/18}{T}^{2/9} \Big\{ 
    3\left[ {T}^{-1/3} - ({t}_{eq})^{-1/3} \right]
    \nonumber \\ & & \phantom{C_2 G^{1/6}{T}^{2/9} \Big\{ }
    + 8 {T}^{-11/24} \left[ {T}^{1/8} - ({t}_{in})^{1/8} 
    \right] \Big\} . \nonumber \\
\eea
This regime of $T$ contains the configuration which maximizes
$\rho_\gamma(t_{eq})/\rho_b(t_{eq})$.
We denote by $T_{\rm max}$ the value of $T$ which maximizes
$\rho_\gamma(t_{eq})/\rho_b(t_{eq})$; this value is found 
numerically to be
\be
  \frac{T_{\rm max}}{\sqrt{G}} = 6.6 \times 10^{50} .
\ee
Note that this value is determined by $t_{in}$ and $t_{eq}$ alone;
it does not depend on any other parameters of our model, all of which
have been sequestered into the constant $C_2$.
In terms of redshift, $T_{\rm max}$ translates to $z_{\rm max} = 7.4 \times 10^5$;
from this we see that $T_{\rm max}$ is quite close to the initial time.
The maximum value of the non-thermal photon energy density is thus
\be \label{rhomax}
  \frac{\rho_\gamma(t_{eq})}{\rho_b(t_{eq})}\bigg|_{T = T_{\rm max}}
  = C_2 (1.3 \times 10^{-5}) .
\ee

We conclude this section by discussing the numerical value of $C_2$.
This constant is a conglomeration of the many constants of our model;
explicitly, $C_2$ may be written
\be
 C_2 \approx 23 \,N f \alpha^{5/2} \beta^{-11/6} \gamma^{-17/18}
 \kappa^{1/3} \lambda^{-1/18} .
 \ee
In the text above we have given the order-of-magnitude estimates 
for these parameters as
\be
 N = \beta = \gamma = 10, \quad
 \alpha = f = 0.1, \quad
 \lambda = \kappa = 1 ,
\ee
which yields the value
\be \label{C2}
  C_2 = 1.2 \times 10^{-4\pm 2} .
\ee
We include in
$C_2$ an uncertainty of two orders of magnitude. This
is due mainly to the uncertainty in the parameters $\alpha$
and $N$ which need to be determined by numerical cosmic
string simulations.
Combining (\ref{rhomax}) and (\ref{C2}) we obtain the approximate 
maximum value of the photon density
\be
  \frac{\rho_\gamma(t_{eq})}{\rho_b(t_{eq})}\bigg|_{\rm max}
  = 1.6 \times 10^{-9 \pm 2} .
\ee
The value of the string tension associated to this maximal value is
\be
  \mu G |_{\rm max} \sim 2.2 \times 10^{-13} .
\ee

In the next section we turn to the implications of these results 
for the magnitude of CMB spectral distortions.

\section{Current and Future Constraints on the Cosmic String Tension}

Non-thermal injection of photons in the redshift interval between 
$z \sim 3 \times 10^6$
and recombination leads to spectral distortions in the CMB (see e.g. \cite{Chluba} for 
a recent comprehensive overview, and \cite{Wright} for an earlier paper). 
For $3 \times 10^6 > z > 10^5$, the energy injection
leads to a Bose-Einstein distribution modified by a chemical potential
($\mu$ distortion). The chemical potential $\mu$ generated by energy
injection is given by
\be
\mu \, = \, \frac{1}{0.7} \frac{\delta U}{U} \, ,
\ee
where $\delta U$ is the energy density in injected photons,
and $U$ is the background photon density. In our case,
$\delta U = \rho_{\gamma}$, the energy density from
cusp evaporation computed in the previous section, and
$U = \rho_b$.

For redshifts in the range $10^5 > z > 3.6 \times 10^3$ (the lower
bound being the redshift of matter-radiation equality), the energy
injection produces a Comptonized spectrum characterized by
a y-distortion, with the y-parameter given by
\be
y \, = \, \frac{1}{4} \frac{\delta U}{U} \, .
\ee

The best limits on spectral distortions still come from the COBE
experiment, and are \cite{Fixsen}
\be
|\mu| \, < \, 9 \times 10^{-5}  \, ,
\ee
and
\be
|y| \, < \, 15 \times 10^{-6} \, ,
\ee
respectively, both at 95\% confidence level. 
These bounds give essentially the same order of magnitude
constraint on the injected photon energy density, which
is thus independent of time:
\be
\frac{\rho_{\gamma}(t)}{\rho_b(t)} \, < \, 10^{-5} \, .
\ee
The proposed PIXIE experiment \cite{PIXIE} will improve 
the constraint by four orders of magnitude to
\be 
\frac{\rho_{\gamma}(t)}{\rho_b(t)} \, < \, 10^{-9} \, .
\ee

In the previous section we found that for the optimal value of $T$
the injected photon density is $1.6 \times 10^{-9\pm 2}$. 
This value is too small to be constrained
by COBE data; however, it does potentially lie within the sensitivity 
of PIXIE. Taking a large value for $C_2$ (i.e., considering
the $+2$ in (\ref{C2})), our model predicts that PIXIE would
detect CMB distortions due to cosmic string loops with string
tension in the range
\be \label{range}
   2.2 \times 10^{-15} < \mu G < 9.4 \times 10^{-12} .
\ee
Smaller values of $C_2$ result in a smaller range, and 
sufficiently small values of $C_2$ (e.g., considering the $-2$ in (\ref{C2}))
result in CMB distortion levels below the sensitivity of PIXIE.


\section{Conclusions and Discussion}

Cusp annihilation is a mechanism by which cosmic strings lose
energy to photons. 
Within our model of cosmic string evolution, cusp
annihilation is the dominant decay channel at early times.
After the cross-over time $T$, cusp annihilation gives way to 
gravitational radiation as the dominant decay channel.
The photons produced by cusp annihilation are out-of-equilibrium
and can lead to spectral distortions of the CMB. 
We have computed the energy density of this injected photon flux as 
a function of the cross-over time $T$ (equivalently, as a function of
the string tension $\mu G$). We find that the resulting CMB distortion
is largest for a cross-over time $T$ at redshift $\sim 7 \times 10^5$,
corresponding to a string tension $\mu G \sim 2 \times 10^{-13}$.

Current limits on $\mu$ and $y$ distortions 
of the CMB are not sensitive enough to constrain cosmic string models. 
However, our analysis shows that the planned 
PIXIE mission will have the sensitivity to rule out (or confirm) 
string tensions in the range of $10^{-15} \lesssim G \mu \lesssim  10^{-12}$ 
which corresponds to symmetry breaking scales $\eta$ of 
$10^{11} {\rm{GeV}} \lesssim \eta \lesssim 10^{13} {\rm{GeV}}$.

An important caveat is that our analysis does not take
the gravitational back-reaction of cusps into account.
Back-reaction may prevent cusps of the length predicted by the 
Nambu-Goto approximation
to develop. In this case, less photons
would be produced and the constraints on the cosmic string
tension would be weaker.

Note added: After the initial submission of this work we became
aware that spectral distortions from cosmic string loop cusp decay
were already considered in Section 4.2 of \cite{Long2} (which is based
in part on \cite{Long1}). Our qualitative results agree, although
the quantitative details differ a bit due to different assumptions made
about the structure of a cusp. This difference highlights the importance
of improved studies of back-reaction effects on cusps

\acknowledgements{RB is supported by an NSERC Discovery Grant, by funds 
from the Canada Research Chair program and by a Simons Foundation Fellowship.
He thanks the Institute for Theoretical Studies of the ETH Z\"urich for 
hospitality during the completion of this project.
IM is supported in part by fellowships
from the Institute of Particle Physics, a
Trottier postdoctoral fellowship, and funds
from NSERC Discovery grants.}


\begin{thebibliography}{99}

\bibitem{Kibble1}

T.~W.~B.~Kibble,
  ``Topology of Cosmic Domains and Strings,''
  J.\ Phys.\ A {\bf 9}, 1387 (1976).

\bibitem{Kibble2}
T.~W.~B.~Kibble,
  ``Phase Transitions In The Early Universe,''
  Acta Phys.\ Polon.\  B {\bf 13}, 723 (1982);\\
  T.~W.~B.~Kibble,
  ``Some Implications Of A Cosmological Phase Transition,''
  Phys.\ Rept.\  {\bf 67}, 183 (1980).

\bibitem{Vil}
A. Vilenkin and E.P.S. Shellard, 
\textit{Cosmic Strings and other Topological Defects} 
(Cambridge Univ. Press, Cambridge, 1994).

\bibitem{HK}
M.~B.~Hindmarsh and T.~W.~B.~Kibble,
  ``Cosmic strings,''
  Rept.\ Prog.\ Phys.\  {\bf 58}, 477 (1995)
  [arXiv:hep-ph/9411342].

\bibitem{RHB1}
R.~H.~Brandenberger,
  ``Topological defects and structure formation,''
  Int.\ J.\ Mod.\ Phys.\ A {\bf 9}, 2117 (1994)
  [arXiv:astro-ph/9310041].

\bibitem{RHB2}
  R.~H.~Brandenberger,
  ``Searching for Cosmic Strings in New Observational Windows,''
  Nucl.\ Phys.\ Proc.\ Suppl.\  {\bf 246-247}, 45 (2014)
  [arXiv:1301.2856 [astro-ph.CO]].
  
\bibitem{CSold}
N.~Turok and R.~H.~Brandenberger,
  ``Cosmic Strings And The Formation Of Galaxies And Clusters Of Galaxies,''
  Phys.\ Rev.\ D {\bf 33}, 2175 (1986);\\
H. Sato, ``Galaxy Formation by Cosmic Strings,''
  Prog. Theor. Phys.\  {\bf 75}, 1342 (1986);\\
A. Stebbins, ``Cosmic Strings and Cold Matter'',
  Ap. J. (Lett.) {\bf 303}, L21 (1986).

\bibitem{Boomerang}
P.~D.~Mauskopf {\it et al.}  [Boomerang Collaboration],
  ``Measurement of a Peak in the Cosmic Microwave Background Power Spectrum
  from the North American test flight of BOOMERANG,''
  Astrophys.\ J.\  {\bf 536}, L59 (2000)
  [arXiv:astro-ph/9911444].

\bibitem{Dvorkin}
C.~Dvorkin, M.~Wyman and W.~Hu,
  ``Cosmic String constraints from WMAP and the South Pole Telescope,''
  Phys.\ Rev.\ D {\bf 84}, 123519 (2011)
  [arXiv:1109.4947 [astro-ph.CO]].

\bibitem{Planck}
  P.~A.~R.~Ade {\it et al.}  [Planck Collaboration],
  ``Planck 2013 results. XXV. Searches for cosmic strings and other topological defects,''
  Astron.\ Astrophys.\  {\bf 571}, A25 (2014)
  [arXiv:1303.5085 [astro-ph.CO]].

\bibitem{older}
L.~Pogosian, S.~H.~H.~Tye, I.~Wasserman and M.~Wyman,
  ``Observational constraints on cosmic string production during brane
  inflation,''
  Phys.\ Rev.\  D {\bf 68}, 023506 (2003)
  [Erratum-ibid.\  D {\bf 73}, 089904 (2006)]
  [arXiv:hep-th/0304188];\\
M.~Wyman, L.~Pogosian and I.~Wasserman,
  ``Bounds on cosmic strings from WMAP and SDSS,''
  Phys.\ Rev.\  D {\bf 72}, 023513 (2005)
  [Erratum-ibid.\  D {\bf 73}, 089905 (2006)]
  [arXiv:astro-ph/0503364];\\
A.~A.~Fraisse,
  ``Limits on Defects Formation and Hybrid Inflationary Models with
  Three-Year WMAP Observations,''
  JCAP {\bf 0703}, 008 (2007)
  [arXiv:astro-ph/0603589];\\
U.~Seljak, A.~Slosar and P.~McDonald,
  ``Cosmological parameters from combining the Lyman-alpha forest with CMB,
  galaxy clustering and SN constraints,''
  JCAP {\bf 0610}, 014 (2006)
  [arXiv:astro-ph/0604335];\\
  R.~A.~Battye, B.~Garbrecht and A.~Moss,
  ``Constraints on supersymmetric models of hybrid inflation,''
  JCAP {\bf 0609}, 007 (2006)
  [arXiv:astro-ph/0607339];\\
R.~A.~Battye, B.~Garbrecht, A.~Moss and H.~Stoica,
  ``Constraints on Brane Inflation and Cosmic Strings,''
  JCAP {\bf 0801}, 020 (2008)
  [arXiv:0710.1541 [astro-ph]];\\
N.~Bevis, M.~Hindmarsh, M.~Kunz and J.~Urrestilla,
  ``CMB power spectrum contribution from cosmic strings using  field-evolution
  simulations of the Abelian Higgs model,''
  Phys.\ Rev.\  D {\bf 75}, 065015 (2007)
  [arXiv:astro-ph/0605018];\\
N.~Bevis, M.~Hindmarsh, M.~Kunz and J.~Urrestilla,
  ``Fitting CMB data with cosmic strings and inflation,''
  Phys.\ Rev.\ Lett.\  {\bf 100}, 021301 (2008)
  [astro-ph/0702223 [ASTRO-PH]];\\
R.~Battye and A.~Moss,
  ``Updated constraints on the cosmic string tension,''
  Phys.\ Rev.\ D {\bf 82}, 023521 (2010)
  [arXiv:1005.0479 [astro-ph.CO]].

\bibitem{RHB3}
 R.~H.~Brandenberger,
  ``Probing Particle Physics from Top Down with Cosmic Strings,''
  Universe {\bf 1}, no. 4, 6 (2013)
  [arXiv:1401.4619 [astro-ph.CO]].

\bibitem{CSsimuls}

A.~Albrecht and N.~Turok,
  ``Evolution Of Cosmic Strings,''
  Phys.\ Rev.\ Lett.\  {\bf 54}, 1868 (1985);\\
D.~P.~Bennett and F.~R.~Bouchet,
  ``Evidence For A Scaling Solution In Cosmic String Evolution,''
  Phys.\ Rev.\ Lett.\  {\bf 60}, 257 (1988);\\
B.~Allen and E.~P.~S.~Shellard,
  ``Cosmic String Evolution: A Numerical Simulation,''
  Phys.\ Rev.\ Lett.\  {\bf 64}, 119 (1990);\\
C.~Ringeval, M.~Sakellariadou and F.~Bouchet,
  ``Cosmological evolution of cosmic string loops,''
  JCAP {\bf 0702}, 023 (2007)
  [arXiv:astro-ph/0511646];\\
V.~Vanchurin, K.~D.~Olum and A.~Vilenkin,
  ``Scaling of cosmic string loops,''
  Phys.\ Rev.\  D {\bf 74}, 063527 (2006)
  [arXiv:gr-qc/0511159];\\
L.~Lorenz, C.~Ringeval and M.~Sakellariadou,
  ``Cosmic string loop distribution on all length scales and at any redshift,''
  JCAP {\bf 1010}, 003 (2010)
  [arXiv:1006.0931 [astro-ph.CO]];\\
J.~J.~Blanco-Pillado, K.~D.~Olum and B.~Shlaer,
  ``Large parallel cosmic string simulations: New results on loop production,''
  Phys.\ Rev.\ D {\bf 83}, 083514 (2011)
  [arXiv:1101.5173 [astro-ph.CO]];\\
J.~J.~Blanco-Pillado, K.~D.~Olum and B.~Shlaer,
  Phys.\ Rev.\ D {\bf 89}, no. 2, 023512 (2014)
  [arXiv:1309.6637 [astro-ph.CO]].

\bibitem{grav}

T.~Vachaspati and A.~Vilenkin,
  ``Gravitational Radiation from Cosmic Strings,''
  Phys.\ Rev.\ D {\bf 31}, 3052 (1985);\\
R. L. Davis, 
``Nucleosynthesis Problems for String Models of Galaxy Formation",
Phys. Lett. {\bf B 161}, 285 (1985).

\bibitem{KS}

N.~Kaiser and A.~Stebbins,
 ``Microwave Anisotropy Due To Cosmic Strings,''
  Nature {\bf 310}, 391 (1984).
\bibitem{Holder1}
R.~J.~Danos, R.~H.~Brandenberger and G.~Holder,
  ``A Signature of Cosmic Strings Wakes in the CMB Polarization,''
  Phys.\ Rev.\ D {\bf 82}, 023513 (2010)
  [arXiv:1003.0905 [astro-ph.CO]].

\bibitem{Holder2}
R.~H.~Brandenberger, R.~J.~Danos, O.~F.~Hernandez and G.~P.~Holder,
  ``The 21 cm Signature of Cosmic String Wakes,''
  JCAP {\bf 1012}, 028 (2010)
  [arXiv:1006.2514 [astro-ph.CO]].

\bibitem{Lin}
A.~Barton, R.~H.~Brandenberger and L.~Lin,
  ``Cosmic Strings and the Origin of Globular Clusters,''
  JCAP {\bf 1506}, no. 06, 022 (2015)
  [arXiv:1502.07301 [astro-ph.CO]];\\
L.~Lin, S.~Yamanouchi and R.~Brandenberger,
  ``Effects of Cosmic String Velocities and the Origin of Globular Clusters,''
  arXiv:1508.02784 [astro-ph.CO].

\bibitem{SMBH}
S.~F.~Bramberger, R.~H.~Brandenberger, P.~Jreidini and J.~Quintin,
  ``Cosmic String Loops as the Seeds of Super-Massive Black Holes,''
  JCAP {\bf 1506}, no. 06, 007 (2015)
  [arXiv:1503.02317 [astro-ph.CO]].

\bibitem{Maddy}
M.~Anthonisen, R.~Brandenberger and P.~Scott,
  ``Constraints on cosmic strings from ultracompact minihalos,''
  Phys.\ Rev.\ D {\bf 92}, no. 2, 023521 (2015)
  [arXiv:1504.01410 [astro-ph.CO]].

\bibitem{KT}
T.~W.~B.~Kibble and N.~Turok,
  ``Selfintersection of Cosmic Strings,''
  Phys.\ Lett.\ B {\bf 116}, 141 (1982).

\bibitem{Olum}
J.~J.~Blanco-Pillado, K.~D.~Olum and B.~Shlaer,
  ``Cosmic string loop shapes,''
  arXiv:1508.02693 [astro-ph.CO].

\bibitem{UHECR}
P.~Bhattacharjee,
  ``Cosmic Strings and Ultrahigh-Energy Cosmic Rays,''
  Phys.\ Rev.\ D {\bf 40}, 3968 (1989);\\
J.~H.~MacGibbon and R.~H.~Brandenberger,
  ``High-energy neutrino flux from ordinary cosmic strings,''
  Nucl.\ Phys.\ B {\bf 331}, 153 (1990);\\
P.~Bhattacharjee and N.~C.~Rana,
  ``Ultrahigh-energy Particle Flux From Cosmic Strings,''
  Phys.\ Lett.\ B {\bf 246}, 365 (1990);\\
P.~Bhattacharjee, C.~T.~Hill and D.~N.~Schramm,
  ``Grand unified theories, topological defects and ultrahigh-energy cosmic rays,''
  Phys.\ Rev.\ Lett.\  {\bf 69}, 567 (1992);\\
J.~H.~MacGibbon and R.~H.~Brandenberger,
  ``Gamma-ray signatures from ordinary cosmic strings,''
  Phys.\ Rev.\ D {\bf 47}, 2283 (1993)
  [astro-ph/9206003];\\
G.~Sigl, D.~N.~Schramm and P.~Bhattacharjee,
  ``On the origin of highest energy cosmic rays,''
  Astropart.\ Phys.\  {\bf 2}, 401 (1994)
  [astro-ph/9403039];\\
U.~F.~Wichoski, J.~H.~MacGibbon and R.~H.~Brandenberger,
  ``High-energy neutrinos, photons and cosmic ray fluxes from VHS cosmic strings,''
  Phys.\ Rev.\ D {\bf 65}, 063005 (2002)
  [hep-ph/9805419].


\bibitem{Carter}
R.~H.~Brandenberger, B.~Carter and A.~C.~Davis,
 ``Microwave background constraints on decaying defects,''
 Phys.\ Lett.\ B {\bf 534}, 1 (2002)
 [hep-ph/0202168].

\bibitem{Tashiro1}
H.~Tashiro, E.~Sabancilar and T.~Vachaspati,
 ``CMB Distortions from Damping of Acoustic Waves Produced by Cosmic Strings,''
 JCAP {\bf 1308}, 035 (2013)
 [arXiv:1212.3283 [astro-ph.CO]].

\bibitem{Tashiro2}
H.~Tashiro, E.~Sabancilar and T.~Vachaspati,
 ``CMB Distortions from Superconducting Cosmic Strings,''
 Phys.\ Rev.\ D {\bf 85}, 103522 (2012)
 [arXiv:1202.2474 [astro-ph.CO]].

\bibitem{CSBH}
J.~H.~MacGibbon, R.~H.~Brandenberger and U.~F.~Wichoski,
 ``Limits on black hole formation from cosmic string loops,''
 Phys.\ Rev.\ D {\bf 57}, 2158 (1998)
 [astro-ph/9707146].

\bibitem{Rachel}
R.~Jeannerot, X.~Zhang and R.~H.~Brandenberger,
 ``Non-thermal production of neutralino cold dark matter from cosmic string decays,''
 JHEP {\bf 9912}, 003 (1999)
 [hep-ph/9901357].

\bibitem{cusp}
D.~N.~Spergel, T.~Piran and J.~Goodman,
  ``Dynamics of Superconducting Cosmic Strings,''
  Nucl.\ Phys.\ B {\bf 291}, 847 (1987).

\bibitem{RHB4}
R.~H.~Brandenberger,
  ``On the Decay of Cosmic String Loops,''
  Nucl.\ Phys.\ B {\bf 293}, 812 (1987);\\
R.~H.~Brandenberger and A.~Matheson,
  ``Cosmic String Decay,''
  Mod.\ Phys.\ Lett.\ A {\bf 2}, 461 (1987).

\bibitem{Chluba}
J.~Chluba,
  ``Science with CMB spectral distortions,''
  arXiv:1405.6938 [astro-ph.CO];\\
 J. Chluba,
  ``Spectral Distortions of the Cosmic Microwave Background'',
  PhD thesis, LMU, 2005.
  
\bibitem{Wright}
E.~L.~Wright {\it et al.},
  ``Interpretation of the COBE FIRAS spectrum,''
  Astrophys.\ J.\  {\bf 420}, 450 (1994).

\bibitem{Fixsen}
D.~J.~Fixsen, E.~S.~Cheng, J.~M.~Gales, J.~C.~Mather, R.~A.~Shafer and E.~L.~Wright,
  ``The Cosmic Microwave Background spectrum from the full COBE FIRAS data set,''
  Astrophys.\ J.\  {\bf 473}, 576 (1996)
  [astro-ph/9605054].
 

\bibitem{PIXIE}
A.~Kogut {\it et al.},
  ``The Primordial Inflation Explorer (PIXIE): A Nulling Polarimeter for Cosmic Microwave Background Observations,''
  JCAP {\bf 1107}, 025 (2011)
  [arXiv:1105.2044 [astro-ph.CO]].

\bibitem{Long2}
A.~J.~Long and T.~Vachaspati,
 ``Cosmic Strings in Hidden Sectors: 2. Cosmological and Astrophysical Signatures,''
 JCAP {\bf 1412}, no. 12, 040 (2014)
 [arXiv:1409.6979 [hep-ph]].

\bibitem{Long1}
A.~J.~Long, J.~M.~Hyde and T.~Vachaspati,
 ``Cosmic Strings in Hidden Sectors: 1. Radiation of Standard Model Particles,''
 JCAP {\bf 1409}, no. 09, 030 (2014)
 [arXiv:1405.7679 [hep-ph]].

\end{thebibliography}
\end{document}